\documentclass[useAMS,usenatbib]{mn2e}
\usepackage[dvips]{graphicx}
\title[BT2 water line list]{A high accuracy computed water line list}
\author[R. J. Barber, J. Tennyson, G. J. Harris,
R. N. Tolchenov]{R. J. Barber, J. Tennyson, G. J. Harris,
R. N. Tolchenov \\ Department of Physics and Astronomy, University
College London, Gower Street, WC1E 6BT London, UK}
\begin{document}

\date{Accepted XXXX. Received XXXX; in original form XXXX}

\pagerange{\pageref{firstpage}--\pageref{lastpage}} \pubyear{2005}

\maketitle

\label{firstpage}

\begin{abstract}
A computed list of H$_{2}$$^{16}$O infra-red transition frequencies and
intensities is presented. The list, BT2, was produced using a discrete
variable representation two-step approach for solving the
rotation-vibration nuclear motions. It is the most complete water line
list in existence, comprising over 500 million transitions (65\% more
than any other list) and it is also the most accurate (over 90\% of
all known experimental energy levels are within 0.3 cm$^{-1}$ of the
BT2 values).  Its accuracy has been confirmed by extensive testing
against astronomical and laboratory data.

The line list has been used to identify individual water lines in a
variety of objects including: comets, sunspots, a brown dwarf and the
nova-like object V838 Mon. Comparison of the observed intensities with
those generated by BT2 enables physical values to be derived for these
objects.  The line list can also be used to provide an opacity for
models of the atmospheres of M-dwarf stars and assign previously
unknown water lines in laboratory spectra.

\end{abstract}


\section{Introduction}
Water is the most abundant molecule in the universe after H$_{2}$ and
CO. It is present in many astrophysical environments including the
atmospheres of: M dwarfs (Allard et al.1994), brown dwarfs (Allard et
al. 1996), K and M giants and supergiants (Jennings and Sada 1998;
Ryde et al. 2002; Tsuji 2001) and oxygen-rich AGB stars (Barlow et al.
1996). It occurs in: sun-spots (Wallace et al. 1992; Polyansky et
al. 1997), nova outflows (Banerjee et al. 2005), Mira variables
(Hinkle and Barnes 1979), T Tauri eruptive variables (Shiba et
al. 1993), dark molecular clouds (Gensheimer et al. 1996), young
stellar objects (Carr et al. 2004), comets (Mumma et al. 1996; Dello
Russo et al. 2000), the ISM (Cernicharo 1994), masers (Cheung et
al. 1969; Gonz\'{a}lez-Alphonso et al. 1995) and planetary
atmospheres. An accurate water line list is thus essential for
interpreting spectra from all of these sources and in modelling
stellar atmospheres at temperatures up to 4,000 K.

The importance of water has given rise to many laboratory
investigations of its spectrum. Ludwig 1971; Camy-Peyret et al. 1977;
Bernath 1996 all investigated hot water line positions. However,
techincal problems and the huge number of transitions (many of which
appear blended) mean that only in the region of 80,000 (out of a total
of more than a billion) transitions are known experimentally and there
are few hot water lines for which intensities have been determined.

The spectrum of water, which extends over a wide wavelength range
from millimetre to near ultra-violet, is due to quantised changes in
the rotation-vibration energy of the atomic nuclei moving in the
electronic potential well. Essentially, the water molecule is only
able to absorb or emit in its ground electronic state as the energy of
the first stable excited electronic state is above the dissociation
energy. In practice, some emissions do occur from short-lived excited
electronic states. H$_{2}$O electronic transitions from diffuse
interstellar clouds provide an example (Smith et al. 1981), but these
transitions, which occur in the ultra-violet, can be disregarded in
almost all other situations.

Water is a triatomic asymmetric top molecule. Its rotation-vibration
spectrum is more complicated than those of most other triatomic
molecules. In common with all non-linear triatomics, H$_{2}$O has six
degrees of internal freedom (three of rotation and three of
vibration). However, the lightness of the hydrogen atoms means that
the rotation constants are large, and this gives rise to an open
spectrum that extends over a wide frequency range. Moreover, the
`floppy' nature of the molecule means that the movement of the
hydrogen atoms is generally anharmonic and consequently transitions
involving changes of more than one vibrational quanta often
occur. Also, since many of the vibrational frequencies are nearly
resonant with other frequencies, it is common for vibrational bands to
overlap and for states to interact in ways that cannot easily be
predicted by perturbation theory but are amenable to a variational
approach.

The importance of the H$_{2}$O molecule in astronomy and the
complexity of its spectrum have created a great deal of interest in
the possibility of generating the spectrum synthetically. Previous
synthetic line lists include: MT (Miller et al. 1994), VT1 (Viti et
al. 1997) VT2 (Viti Ph.D.thesis 1997), PS, otherwise called AMES,
(Partridge \& Schwenke 1997) and SCAN (J{\o}rgensen et al. 2001). All
previous synthetic line lists have suffered from a number of problems
that are discussed below. The most successful attempts have employed
similar variational nuclear motion procedures to that used in
producing the BT2 water line list. However, for reasons detailed in
the next section, none of the earlier lists is considered to be
satisfactory. We have addressed these problems and consequently the
BT2 line list is an accurate tool for astronomers working in a variety
of fields.

\section{Background to the calculations}

\subsection{Variational Techniques}

Variational techniques represent the best approach to solving the the
nuclear motion problem (Tennyson 1992), and they are examined in detail in
Ba\v{c}i\'{c} and Light (1989). Here we use a discrete variable
representation (DVR).

In a DVR, the wavefunctions are defined by a complete set of weighted,
orthogonal grid points, each wavefunction having a different set of
weightings. This method is capable of generating accurate solutions,
the accuracy being determined by the the number and appropriateness of the
points. It is efficient for a large number of situations. The DVR
approach has the advantage that the potential matrix elements are
diagonal, and hence are easily evaluated. Even more important is the
fact that the dipoles can be reduced to a similar form (Tennyson et al.
2004).

The current work is not unique in employing a DVR approach to solve
the nuclear motion problem for water (see for example, Viti et
al. 1997). However, improved physics, in the form of a highly accurate
potential energy surface (PES), the methodology embodied in the DVR3D
program suite and increases in computational power have made it
possible to produce a line list that is more complete and more
accurate than any previous list, even those employing similar
methodolodies.

The DVR3D approach generates four rotation-vibration symmetry blocks
for the H$_{2}$O molecule, which we label ee, eo, oe and oo. The first
e/o term is the vibrational basis symmetry `q' and the second, e/o is
the standard quantum number `p', the rotational parity. These
symmetry blocks are not the same as the C$_{2v}$(M) symmetry
blocks $\Gamma_{rv}$: A$_1$, A$_2$, B$_1$, B$_2$, but are related to
them (see standard texts on molecular spectroscopy, such as Bunker
\& Jensen 2005).

The nuclear permutation operation in which the two identical protons
comprising the hydrogen nuclei are interchanged gives rise two two
dicrete states of the molecule: ortho (O) and para (P).The nuclear
spins may couple symmetrically or antisymmetrically.  The
antisymmetric coupling that gives rise to the O form of the H$_{2}$O
molecule is triply degenerate, whilst the symmetric coupling, that
gives rise to the P form of the molecule is undegenerate.

For any J, apart from J=0, there are two ortho and two para symmetry
blocks (in the case of J=0 there is one O and one P block)
There are two possible arrangements, depending on whether J is odd or
even and these are detailed in Table 1.

\begin{table}
\begin{center}
\caption{Symmetry Blocks}
\begin{tabular}{cccccccccc}
\hline

{ }& \multicolumn{4}{c}{\bf J even}&&\multicolumn{4}{c}{\bf J odd}\\
\cline{2-5}\cline{7-10}

\bf q    & e & e & o & o && e & e & o & o\\
\bf p    & e & o & e & o && e & o & e & o\\
\bf O/P  & P & O & O & P && O & P & P & O\\
\bf Code & 1 & 3 & 4 & 2 && 3 & 1 & 2 & 4\\
\hline
\end{tabular}
\end{center}
{\footnotesize q is the vibrational basis symmetry and p is the
rotational parity. These are labelled as either symmetric (e),
or antisymmetric (o) states. O/P = Ortho/Para. The `Code' is the
notation for symmetry used in the Levels File (Table 2).}

\end{table}

The difference in degeneracies of the O and P states impacts on the
partition function of the molecule as well as on line intensity. Also,
the fact that O-P and P-O transitions are forbidden, has spectroscopic
consequences.

\subsection{PES and the energy levels}
The energies of the quantised rotation-vibration states are
the eigenvalue solutions that satisfy the Shr\"{o}dinger equation for
the oxygen and two hydrogen nuclei moving within the electronic
potenial. However, the electronic potential within which the charged
nuclei are moving is itself a function of the actual internuclear
geometry. Therefore, in order to solve for the nuclear motion, it is
necessary to have an accurate model of how the potential varies with
the nuclear geometry. The problem is rendered tractable by adopting
the Born-Oppenheimer approximation which separates the nuclear and
electronic motions and has as its basis the fact that, due to their
lightness, the electrons may be considered to react immediately to any
changes in the nuclear geometry. The most accurate PESs are computed
using an \emph{ab initio} starting point, with the resulting surface
being empirically adjusted to improve the agreement between the
computed energies and experimental data (Partridge and Schwenke
1997).

We used the potential energy surface fit B of Shirin et al. (2003),
which is based on a highly accurate \emph{ab initio} surface with
adjustments for electronic relativistic and adiabatic (also known as
the Born-Oppenheimer diagonal correction) effects and fitted to the
available experimental data.  At the time of writing, this surface is
the most accurate available and it is the single most important factor
affecting the accuracy of our results.

Non-adiabatic corrections to the Born-Oppenheimer approximation are
also important in the case of water (Schwenke 2003). However, a full
theoretical non-adiabatic adjustement to the PES has been shown to
produce no significant benefits compared to simplified approaches
(Tennyson et al., 2002). These authors examine two such approaches:
using separate reduced masses for the vibrational and rotational
motions (this systematically over-corrects bending motions and
under-corrects stretches), and a simplified version of the full
correction which includes only terms that scale with the kinetic
energy terms in $\theta$ and r. The latter approach is preferred and
the ajustment is effected through the DVR3D program rather than
changes to the PES.

\subsection{Transition intensities}
 The intensities of the allowed transitions between rotation-vibration
states are determined by the dipole transition moments for these pairs
of states. They are:
\begin{equation}
\label{Eq:A}
\quad\qquad \qquad \qquad \qquad \langle
\psi^{\prime}|\overline\mu|\psi^{\prime\prime}\rangle
\end{equation}
where $\psi^{\prime}$ and $\psi^{\prime\prime}$ are the wavefunctions
of the two states,and $\overline{\mu}$ is the electronic dipole moment
vector.  Therefore, in order to calculate intensities, a dipole moment
surface (DMS) is required.

The parameter computed by the DVR3D program suite is the Einstein A
coefficient for, A$_{if}$, for each transition. This is the
coefficient of spontaneous emission between the upper and lower
states. It is related to the dipole transition moment for the pair of
states and to J for the upper state. A$_{if}$, is independent of
temperature, relates to a single molecule, has units of s$^{-1}$ and
is given by:

\begin{equation}
\label{Eq:B}
\qquad A_{if} =
\frac{64\pi^{4}}{3c^{3}h}\nu^{3}g_{i}(2J^{\prime\prime}+1)|\langle
\psi^{\prime} | \overline{\mu} | \psi^{\prime\prime} \rangle|^2
\end{equation}
 where prime and double prime represent the upper, i, and lower, f,
 states respectively. 

The quantity usually derived from observation, is the line intensity,
I, which has units of cm/molecule. I is temperature-dependent and in
emission is related to A$_if$ by the expression:

\begin{equation}
\label{Eq:C}
I = \frac{C(2J^{\prime}+1)}{Q_{vr}(T) \nu^{2}
g_i}\exp\left(\frac{-hcE^{\prime\prime}}{kT}\right)\left[1-\exp\left(\frac{-hc\nu}{kT}\right)\right]A_{if}
\end{equation}

where $\nu$ is the frequency in cm$^{-1}$, E$^{\prime\prime}$ is the
energy of the lower ro-vibrational level in cm$^{-1}$. $Q_{vr}$(T) is
the ro-vibrational partition function and is dimensionless, and $g_i$
is the nuclear spin degeneracy and caries only one subscript since
transitions between different nuclear spin states are not allowed
(Miani and Tennyson 2004). Boltzmann's constant, k, has units of
JK$^{-1}$ and the constant C has the value (8$\pi$c)$^{-1}$ =
1.3271x10$^{-12}$ s cm$^{-1}$.

Since the BT2 line list includes A$_{if}$ for each transition, the
above equation enables the line intensities to be computed at any
given temperature.

Unlike the PES, where the most accurate are \emph{ab initio} surfaces
that have been fitted to the available experimental data, in the case
of DMS, limitations on the accuracy of experimental line strengths
means that at the current time the most accurate surfaces are purely
\emph{ab initio} (Lynas-Gray et al. 1995).  Two such DMS were tested
in the intensity part of our calculations. Our initial BT1 line list
was computed using the same PES as BT2 and a preliminary Lynas-Gray et
al. (in preparation) DMS. The line positions in BT1 are identical to those in
BT2 (the same PES having being used for both). However, we observed
that the Einstein A coefficients of the weaker lines generated using
the Lynas-Gray et al. DMS were often too large when compared with
experiment. This part of the computation was therefore repeated using
the DMS of Schwenke and Partridge (2000). The results, which are
contained in BT2, show much better agreement between the computed
strengths of weak lines and experiment. The Einstein A coefficients of
the stronger lines in BT1 and BT2 generally agree to within two
percent and both agree reasonably well with experimental values. In
addition to being superior to the Lynas-Gray et al. surface that we
tested, Schwenke and Partridge's DMS represents a major improvement on
the earlier DMS of Partridge and Schwenke (1997) and from our
analysis, as well from the same authors, is the most accurate in
existence.

\section{Calculating the line list}

The BT1 and BT2 line lists were computed using the DVR3D suite of
programs (Tennyson et al. 2004) on three Sun 5 Microsystem V880
mainframe computers: Enigma and Ra which are clustered using high
speed interconnects, each having 8 processors and 32 Gb of RAM and 432
Gb of disk storage and PSE, which has 24 processors and 96 Gb of RAM
and 1,296 Gb of disk storage at UCL's Hiperspace computing centre.
The final, `DIPOLE' stage of the suite was amenable to parallelisation
with little time penalty. Other parts of the program were run on
single processors, which avoided coding problems and was more
efficient in computer time.  The total number of processor hours
employed in generating the BT1 line list (including the preliminary
convergence testing which is discussed below) was 55,000 hrs and a
further 10,000 hrs. were used in repeating the `DIPOLE' runs to
generate BT2 and in testing the outputs.

DVR3D calculates the bound rotation-vibration energy levels, the
wavefunctions on a grid in three dimensional space, and the dipole
transition strengths of the allowed transitions. The final part of the
DVR3D suite, `SPECTRA' is able to compute temperature-dependent
spectra over any selected frequency range, convolved with either the
natural line width or some other selected profile, such as that given
by the resolving power of a particular spectrometer. DVR3D uses an
exact (within the Born-Oppenheimer approximation) kinetic energy
operator. The program uses a discrete variable notation (DVR) with two
radial and one angular co-ordinate for the nuclear motion problem
(vibrational and rotational motions being treated spearately).

In establishing the working parmeters for our calculations, we aimed
to provide a line list that would be complete and accurate at the
temperatures of late series K-series stars (up to 4,000 K) and at
wavelengths down to 0.8 $\mu$m. Preliminary calculations showed that
in order to achieve this it would be necessary to include all states
lying at energies up to 30,000 cm$^{-1}$ relative to the ground state
of the system. Previous workers (Miller et al. 1994 and
Partridge and Schwenke 1997) selected lower energy cut-offs.

Our cut-off for the total angular momentum was J=50. We calculate that
the highest value of J which has ro-vibrational energies of less than
30,000 cm$^-1$ is 58. However, we also estimate that by terminating
our calculation at J=50 we omit less than 500 levels out of a total of
more than 505 million and none of the omitted levels has an energy
less than 23,490 cm$^{-1}$, with the majority being at energies above
28,000 cm$^{-1}$. Consequently, even at a temperature of 4,000 K these
missing levels contribute less than 0.02\% of the total partition
function. Moreover by omitting Js above 50 we saved in the region of
8,000 processor hours.

In order to generate accurate eigenvalues, it is essential that all
calculations are fully converged within the limitations imposed by
computing power and time. The lack of convergence in earlier lists has
already been noted (Polyansky et al. 1997).

DVR3D has the option of using Jacobi or Radau co-ordinates. The latter
was selected as being more appropriate for water. Our Radau grid is
defined in terms of two radial co-ordinates, r$_{1}$ and r$_{2}$, each
with 28 points, and one angular co-ordinate, $\theta$ having 44
points. These numbers are determined by convergence testing at
J$=$20. The eigenvalue solutions are particularly sensitive to the
number of radial points, r$_n$. However, the computation time rises very
rapidly as this number is increased, so the selection of r$_n$
represents a compromise and is the principal factor impacting on
convergence.

The angular grid points are arrived at using associated Legendre
polynomials for the underlying basis sets, whilst the radial grid
points are set up using Morse oscillator-like underlying basis sets,
which are defined in terms of parameters, r$_{e}$, $\omega_{e}$ and
D$_{e}$. These must be entered into `DVR3DRJZ', the first module of
the DVR suite. The parameters have physical counterparts which are
respectively: the equilibrium bond length, harmonic frequency and
dissociation energy of the water molecule. However, the parameters in
the Morse oscillator function differ from the physical values and must
be determined by empirical testing. 

Although in principle, DVR3D is not strictly a variational method, in
practice it is found that the Variational Principle does apply; this
fact is employed in obtaining values for r$_{e}$, $\omega_{e}$ and
D$_{e}$. The method is to alter the three parameters in a systematic
manner until DVR3DRJZ generates a set of pure vibrational states (the
J=0, para states), the sum of whose energies is a minimum (251
eigenvalues were employed). We conducted this part of the process
manually.

The investigation was complicated by the existence of local minima,
which are not global minima for the three variables. The set of
parameters giving this result was: r$_{e}$ = 2.05, $\omega_{e}$ = 0.008
and D$_{e}$ = 0.20 all in atomic units (the equivalent dissociation
energy is 43,895 cm$^{-1}$). A good choice of these parameters (which
are a function of the system and of the particular energy range that
is of interest) is fundamental to the accuracy of the ensuing
calculations. The eigenvalues are particluarly sensitive to r$_{e}$
and $\omega_{e}$ and it was observed that differences of as little as
0.05 in the first of these two parameters and 0.002 in the second
could affect the energy levels by in the region of 0.01cm$^{-1}$ for
states with energies in the region of 10,000 cm$^{-1}$. Larger
deviations in these basis functions from their determined optimum
values produced correspondingly greater errors and in the case of
states with energies over 20,000 cm$^{-1}$, a bad choice of parameters
could easily result errors in the individual levels in excess of
20 cm$^{-1}$. Consequently considerable time was spent in determining
the values of the Morse oscillator-like basis set.

Two other inputs are required by the vibrational module,
DVR3DRJZ. These are: the maximum size of the intemediate Hamiltonian,
which we chose after testing as 2500 and the number of eigenvectors to
be saved for use in the rotational module, `ROTLEV3B', the optimum
value for which was found by testing to be 700.
 
ROTLEV3B also requies one variable to be determined; this is IBASS,
the size of the Hamiltonian in the rotation module. IBASS varies with
J. Its value was established by convergence testing at J=20 as being
$530\times(J+1-p)$, where p is the rotational parity and has the value
0 for even parity states and 1 for odd parity states. Hamiltonians of
this size are expensive in terms of computing time at high
Js. Nevertheless, it was easy to demonstrate that lower values of
IBASS produce results that are not converged. This is significant, as
earlier workers using similar DVR techniques have used lower IBASS
values. Viti et al. 1997, for example, used $200\times(J+1-p)$, and PS
used an even lower effective number, particularly at high J.

In selecting the various parameters referred to above, we regularly
tested for convergence. We estimate that our choice of r$_{n}$=28
accounted for approximately half of our total convergence error, which
we estimated as being less than 0.01 cm$^{-1}$ at 10,000 cm$^{-1}$ and
in the region of 0.02 cm$^{-1}$ at 20,000 cm$^{-1}$.

\section{Results}
The BT2 water line list is available electronically in compressed form
at: \emph {ftp://cdsarc.u-strasbg.fr/cats/VI/119} The data are in two
parts. The first, the `` Levels File'' is a list of 221,097 energy
levels, ordered by J and symmetry block. About 25,000 of these energy
levels have been labelled with the appropriate angular momentum
(J,$K_a,K_c$) and vibrational ($\nu_1$,$\nu_2$,$\nu_3$) quantum
numbers. An extract from the Levels File with an explanation of the
contents of each of the 11 columns in the file is given in Table 2.

The second part of BT2 is the ``Transitions File''. This has
505,806,202 entries. Each transition references upper and lower energy
levels in the Levels File and gives the Einstein $A_{if}$ coefficient
for the transition. An extract from the Transitions File is given in
Table 3.

In uncompressed form the BT2 Transitions File is 12.6 Gb of
data. Therefore, in order to facilitate use of the list, the
transitions have been ordered by frequency and separated into 16
smaller files, each representing a specific frequency range.

In addition to the files containing the actual line list, the
Strasbourg site contains a Fortran program, spectra-BT2.f90, that will
enable users to generate emission or absorption spectra from BT2 by
specifying various parameters including: temperature, frequency range,
cut-off intensity and line width. There is also a facility to generate
spectra with full ro-vibrational assignments if required. The method
of using spectra-BT2.f90 is detailed in a `readme-spectra' file and
there are also examples of a job file and an output file.

\begin{table}
\caption{Extract from the BT2 Levels File} 
%the heading
\begin{tabular}{|cccrccccccc|}
%c is centred, r is right justified, l is left justified
\hline

 A    & B & C & D & E         & F & G & H & I & J & K \\[0.5ex]
%the last command leaves an additional space
\hline
%horizontal line
 2284 & 2 & 2 & 5 & 3885.718672 & 0 & 0 & 1 & 2 & 2 & 1 \\
 2285 & 2 & 2 & 6 & 4777.145956 & 0 & 3 & 0 & 2 & 1 & 1 \\
 2286 & 2 & 2 & 7 & 5332.258064 & 1 & 1 & 0 & 2 & 1 & 1 \\
 2287 & 2 & 2 & 8 & 5472.371851 & 0 & 1 & 1 & 2 & 2 & 1 \\
 2288 & 2 & 2 & 9 & 6254.694085 & 0 & 4 & 0 & 2 & 1 & 1 \\
 2289 & 2 & 2 &10 & 6876.917089 & 1 & 2 & 0 & 2 & 1 & 1 \\
 2290 & 2 & 2 &11 & 7027.396535 & 0 & 2 & 1 & 2 & 2 & 1 \\
 2291 & 2 & 2 &12 & 7293.201639 & 2 & 0 & 0 & 2 & 1 & 1 \\
 2292 & 2 & 2 &13 & 7376.617020 & 1 & 0 & 1 & 2 & 2 & 1 \\
 2293 & 2 & 2 &14 & 7536.864373 & 0 & 0 & 2 & 2 & 1 & 1 \\
\hline

\end{tabular}
{\footnotesize
%indicates that the footnote is to be normal size text
A: Row in file, B: J, C: Symmetry (1-4: see Table 1), D: Row in block,
E: $\nu$ in cm$^{-1}$ F, G, H: $\nu_1$, $\nu_2$, $\nu_3$. I, J, K: J,
K$_a$,K$_c$. }
%note the open and close curley brackets
\end{table}

\begin{table}
\begin{center}
%centres the table (need to end lower down)
\caption{Extract from BT2 Transitions File} 
\begin{tabular}{|ccc|}
\hline

 A    & B & C  \\[0.5ex]
\hline
 1000 & 239 & 9.671E+01 \\
 1001 & 239 & 1.874E+00 \\
 1002 & 239 & 4.894E-03 \\
 1003 & 239 & 1.140E-04 \\
 1004 & 239 & 1.707E-02 \\
 1005 & 239 & 8.473E-08 \\
 1006 & 239 & 6.535E-04 \\
 1007 & 239 & 7.157E+00 \\
 1008 & 239 & 6.403E-06 \\
 1009 & 239 & 9.861E-05 \\
\hline

\end{tabular} \\
{\footnotesize A, B: Row numbers in the Levels File
(upper and lower levels are not identified as the program tests for
these). C: A$_{if}$ (s$^{-1}$).}
\end{center}
\end{table}

\begin{table}
\begin{center}
\caption{Comparison of BT2 and PS (Partridge \& Schwenke 1997) with 14,889 experimentally-determined energy levels}
\begin{tabular}{|ccc|}
\hline

Within & BT2 & PS \\
cm$^{-1}$ & \% & \% \\
\hline
0.1 & 48.7 & 59.2 \\
0.3 & 91.4 & 85.6 \\
1.0 & 99.2 & 92.6 \\
3.0 & 99.9 & 96.5 \\
5.0 & 100  & 97.0 \\
10.0& 100  & 98.1 \\
\hline
\end{tabular} \\

\end{center}
\end{table}

\subsection{Comparing the BT2 and Partridge \& Schwenke line lists}
Several water line lists are in regular use by astronomers and the the
most accurate list previously is that of Partridge and Schwenke (PS).
Table 4 compares BT2 and PS energy levels with known experimental
values (Tennyson et al. 2001). It will be seen from this table that
although the PS list is more accurate than BT2 in the cases where
agreement with experiment is better than 0.1 cm$^{-1}$, this is not the
case generally. Specifically, based on a sample of 14,889 levels,
whilst 99.9\% of the BT lines are within 3.0 cm$^{-1}$ of experiment
3.5\% of the PS lines are outside this range. Other line lists (MT,
VT2, SCAN) perform significantly worse than this.

Examining deviations from experiment by energy is even more revealing,
for it is seen that PS is increasingly unreliable above 10,000
cm$^{-1}$, which is the region with the greatest number of transitions.

\begin{table}
\begin{center}
\caption{Distribution of levels in the BT2 and PS (Partridge \&
Schwenke 1997) disageeing with experiment by more than 2 cm$^{-1}$ -
by frequency}
\begin{tabular}{rccc}
\hline

Level Energy & Number & BT2 & PS \\
cm$^{-1}$ & in range & No. & No. \\
\hline
20,000 - 26,300 & 575 & 9 & 334 \\
15,000 - 20,000 & 2,813 & 10 & 105 \\
10,000 - 15,000 & 6,323 & 8 & 58 \\
 7,000 - 10,000 & 3,263 & 3 & 9 \\
$<$ 7,000      & 1,914  & 0 & 0 \\
\hline
Total      & 14,889 & 30 & 506 \\
\hline

\end{tabular} \\
\end{center}
\end{table}

\subsection{Labelling the levels}
As with observed water lines, the production of a synthetic line list
prompts the question of how to assign quantum numbers to the
transitions. The BT2 line list contains over 505 million transitions,
but the DVR3D suite only provides data on J and the symmetry block of
each energy level.  Since each individual line is a transition between
two energy states, the problem reduces to one of labelling the 221,097
energy levels, but this is still a large task.

Many papers have been devoted to assigning quantum numbers to energy
levels that have been deduced from experimental line frequencies (e.g.
Zobov et al. 2000). So far, less than 15,000
experimentally-determined levels have been labelled with their 3
rotational and 3 vibrational quantum numbers, even though much effort
has been expended on the task. 25,870 levels in the BT2 list have been
labelled using methodologies detailed below. BT2 labels 270 of the 416
levels which have J=0 and energies below 30,000 cm$^{-1}$. However,
the proportion labelled is less at higher Js.

The process of labelling BT2 started by identifying particular
energies with experimental levels that have already been
determined. In addition, we have labelled many of the energy levels
that are unknown experimentally, using several different methods. A
large number of vibrational states for J=0 have been labelled by
visual inspection of the nodal structure of the wavefunction as
described in Mussa and Tennyson (1998). This method has the
disadvantage that observations can only easily be made on
two-dimensional sections of the three-dimensional wavefunction and the
procedure can be misleading. In addition, Mussa and Tennyson oberved
that at energies in the region of our 30,000 cm$^{-1}$, a high
proportion of the wavefunctions are irregular with no identifyable
nodal structure. .

A second method involved examining the $A_{if}$ coefficients for pure
rotational transitions between an unlabelled vibrational state and a
known vibrational state; the strongest transitions being those where
standard selection rules are obeyed. We found this method (see
Tolchenov et al 2005) to be useful for J$<$10

A third method that was found to be useful in labelling higher
J states involved the use of an algorithm to identify sets of levels
within the same parity block having the same K$_{a}$, $\nu_{1}$,
$\nu_{2}$, $\nu_{3}$ quantum numbers, but different values of J and
K$_{c}$. The method was originally developed for labelling the energy
levels of the HCN and HNC isomers (Barber et al. 2000). However,
because of the density of the energy levels in water, an extra term
was introduced when labelling the water levels. The algorithm used in
this case was:

\begin{equation}
\label{Eq;6}
E_{J_{n}}\cong 4E_{J_{n-1}} -6E_{J_{n-2}} +4E_{J_{n-3}} -E_{J_{n-4}}
\end{equation}
However, when resonance between levels caused the behaviour pattern to
be erratic, or no E$_{J_{n-4}}$ value existed, the original, simpler
algorithm was used:

\begin{equation}
\label{Eq;7}
E_{J_{n}}\cong 3E_{J_{n-1}} -3E_{J_{n-2}} +E_{J_{n-3}}
\end{equation}
where E$_{J_{n}}$ is the energy of the state in the same symmetry block
having J = n and the same set of K$_{a}$, $\nu_{1}$, $\nu_{2}$,
$\nu_{3}$ quantum numbers.

The results of the labelling exercise are included in the BT2 Level
File. This means that when synthetic spectra are generated many of the
transitions are fully labelled. This feature is useful when generating
synthetic spectra for astronomical or laboratory applications as is
discussed in the next section.

\subsection{Completeness}
If we compare the partition function, Q(T) for water computed at a
particular temperature using the BT and PS line lists with the most
accurately known value at this temperature, it is possible to estimate
the the completeness of the line lists and the amounts of opacity that
are missing in spectra generated by the two line lists at the selected
temperature.

A calculation of the partition function of water at 3,000K using the
221,097 energy levels in BT2 yields a value that is 99.9915 percent of
the Vidler \& Tennyson (2000) value, which indicates that levels
missing from BT2 only contribute about 85 parts in a million to the
partition function of water at this temperature, the reason being that
there is a diminishingly small probability of states above 30,000
cm$^{-1}$ being occupied at this temperature. For comparison, the PS
line list, which has 28,000 cm$^{-1}$ cut-off gives a partition
function at 3,000 K that is only 99.493$\%$ of the Vidler \& Tennyson
value.

Although the exclusion of levels above 30,000 cm$^{-1}$ does not
materially affect the completeness of the BT2 list, it does affect
absorption at shorter wavelengths. If we consider a photon of of
wavelength 1 $\mu$m (energy 10,000 cm$^{-1}$). This is able to be
absorbed by a water molecule in a particular rotation-vibration
eigenstate provided that there is another eigenstate exactly 10,000
cm$^{-1}$ above this lower state into which the molecule may be
excited. It follows that since the BT2 line list has an upper cut-off
of 30,000 cm$^{-1}$, none of the energy levels in the list above
20,000 cm$^{-1}$ are capable of being excited by a 1$\mu$m photon,
since there is no corresponding upper level.

If we examine the extent to which Q(3,000K) computed from BT2, but
excluding all levels above 20,000 cm$^{-1}$, falls short of the Vidler
\& Tennyson value we will have an indication (this is an upper limit
as it takes no account of blending effects) of the opacity that has
been excluded by adopting a 30,000 cm$^{-1}$ cut-off. Performing the
calculation gives a shortfall of 0.83$\%$.

In the case of PS, only energy levels below 18,000 cm$^{-1}$ are able
to absorb a 1$\mu$m photon, and computing Q(3,000K) using only PS
states up to 18,000 cm$^{-1}$ shows that these comprise only
98.37$\%$ of the Vidler value. Hence it will be seen that the opacity
deficit at 1$\mu$m is in the region of twice as great in the PS list
as in BT and the ratio increases at shorter wavelengths.

\section{Sample Applications}
Although BT2 shows good agreement with experimentally known lines
this is not a sufficient test of its accuracy as the PES used to
generate BT2 was fitted to the known experimental data (this is also
true of the PS line list).

The most effective way of checking the accuracy of a line list is to
test its ability to predict or identify previously unknown lines in
astronomical or laboratory spectra. BT2 has been used successfully as
outlined below. It should be noted here that some of the earlier
spectroscopic applications used BT1 as they predate BT2. However, the
line positions of the two lists are identical, and the intensities of
the strong lines are similar for the two lists.

\subsection{Astronomical Spectra}
\subsubsection{$\epsilon$ Indi Ba}
A synthetic spectrum generated at 1,500K using BT1 was able to
reproduce the previously unknown absorption feature observed at 1.554
$\mu$m in the spectrum of the early T dwarf $\epsilon$ Indi Ba,
discovered by Volk et al. (2003), as being a blend of six water lines
with no individual line contributing more than 25$\%$ of intensity
(Smith et al. 2003).

%\begin{figure}
%\includegraphics[width=95mm]{vvsmith.ps}
%\caption{Identification of water features in Brown Dwar\epsilon Indi Ba}
%\end{figure}

\subsubsection{Comet 153P/Ikeya-Zhang (2002 C1)}
BT1 was used to compute the frequencies and Einstein A coefficients of
the 64 transitions (up to J=7) that make up each of the 7 hot bands of
water detected in Comet 153P/Ikeya-Zhang. Dello Russo et al. (2004)
applied these data in determining the rotational temperature of the
comet on three dates.

\subsubsection{Temperatures of comet forming regions in early solar
  nebula}

The hot-band transitions identified by BT1 in Dello Russo et
al. (2004) were classified into ortho and para using the symmetry
information contained within the BT2 energy file (see column C Table
1). Transitions between different nuclear spin parities can be ignored
(Miani and Tennyson 2004). Dello Russo et al. (2005) were able to
deduce the primordial O/P water composition of three comets: C/1999
H1, C/1999 S4 and C/2001 A2 and hence the temperatures of the
different regions of the early solar nebula in which the comets were
formed. The normal O/P ratio is 3:1, but since the lowest ortho level
lies 23.8 cm$^{-1}$ ($\sim$ 34 K) above the ground state (which is a
para state), the O/P ratio is sub-normal at temperatures below $\sim$
50 K. A comparison of the Einstein A coefficients in BT2 with those
actually used in BT1 shows that Dello Russo et al.'s results would
have been the same had the later line list been used.

\subsubsection{Detection of water lines in nova-like object V838 Mon}
Synthetic spectra generated by BT2 were used to identify absorption
features observed in the 1.73 to 1.75 $\mu$m region of the spectrum of
the nova-like object, V838 Monocerotis on five separate dates as being
due to blended water lines. Quantum numbers were assigned to the 17
strong transitions that comprise the 5 absorption features (Banerjee
et al. 2005). Sixteen of the lines were found to be in the (0 0
0)-(0 1 1) band.  This is the first time that individual water lines
have been identified in a nova-like outflow region.

In addition, BT2 was used to compute the theoretical intensities of
the five absorption features as a function of temperature and column
density (assuming LTE). The results indicated that the water features
were arising from a cool $\sim$ 750--900 K region around V838 Mon that
was cooling at a rate of $\sim$ 100 K per year. Column
densities computed for the five dates also showed a reduction with
time.

\subsubsection{Sunspot spectra}
 Following their assignment of high temperature laboratory water lines
 in which extensive use was made of BT2 (see `Laboratory Spectra'
 below), Coheur et al. (2005) revisited the sunspot absorpton spectrum
 in the 9.89--12.95 $\mu$m region of Wallace et al. (1995) which had
 been partially assigned by Polyansky et al. 1997 and Zobov et
 al. 1999. Coheur et al. used their high temperature laboratory
 assignments to identify a substantial number of previously unassigned
 sunspot lines.

\subsection{Laboratory Spectra}
Hot water spectra have been analysed in the laboratory over the last
thirty years. Methods generally involve high resolution Fourier
transform spectrometry of vapour which may be at an elevated
temperature, such as in an oxy-acetylene flame (Camy-Pyret et
al. 1977, Coheur et al. 2005).

The BT2 line list has already been used on a number of occasions to
analyse spectra generated at both high and low temperatures at various
wavelengths and this has has added considerably to the existing
database of experimentally known energy levels and transitions.

Coheur et al. (2005) working with experimental spectra generated at
3,000 K in the 5--20 $\mu$m region labelled about 600 previously
unidentified levels using the BT2 line list. The identification of
these levels was an important factor in their subsequent assignment of
8,500 of the 10,100 lines that they observed in the 0.385-1.081 $\mu$m
region.  Most of the states that were labelled by Coheur et al. were
either high J states having low bending modes or else lower J states
with higher bending modes. These states had defied earlier analysis
because the previous line lists used in earlier work, were unable to
accurately predict the energies of states above 15,000 cm$^{-1}$ or to
treat accurately the high bending mode states, even those below 15,000
cm$^{-1}$.

They note that whilst the agreement between BT2 and
observation is generally within 0.1 cm$^{-1}$ for the energy levels
that they observe for low J states, the disageement can be as great as
0.8 cm$^{-1}$ for some of the very high J states. Nevertheless, they
comment that since for a given vibrational and K$_a$ state the
diffence between the BT2 list and experiment increases smoothly with
J, they are able to use BT2 to predict the positions of unknown higher
J levels with an accuracy of 0.02 cm$^{-1}$. This is considerably less
than the experimentally determined line widths (full widths at half
maximum) which were in the range 0.05--0.10 cm$^{-1}$ due to
broadening at 1 atm. and T $=$ 3000 K.

In a follow-up paper analysing the 2--5$\mu$m region Zobov et al.,
2006 use BT2 to label approximately 700 previously
unidentified ro-vibrational energy levels for water, observed in
laboratory torch spectra.

Tolchenov et al. 2005 analysed long-pathlength room temperature
spectra. They use three separate line lists in their work. However,
only the BT2 list is found to be reliable over all transition
frequencies and in the case of lines having frequncies above 16,000
cm$^{-1}$ it is the only one that they use. In addition, Tolchenov et
al.  encounter difficulties with the labelling adopted in previous
studies, finding, for example, that in some cases different states
have been labelled with the same quantum numbers. They therefore
undertake a systemmatic re-labelling exercise using the BT2
energies. These labels are incorporated into the BT2 list for states
with J $\le$ 9.

Dupr\'{e} et al.(2005) found BT2 similarly reliable for predictions of
long-pathlength room temperature spectra in the near
ultra-violet. They observed 62 R-brach transitions in the 8$\nu$
polyad and were able to determine 36 energy levels, previously unknown
experimentally.

\section{Conclusion}
We present a new synthetic water line list which gives the energies of
221,097 states with cut-offs of J=50 and E=30,000 cm$^{-1}$. 25,870 of
the lower energy levels have been labelled with a full set of three
rotational and three vibrational quantum numbers.  BT2 lists
505,806,202 trasitions. It has been extensively tested against
experimental observations and also compared with other lists. It has
been shown to be the most complete and accurate water line list in
existence.

We make our results freely available in electronic form via
\emph{ftp://cdsarc.u-strasbg.fr/cats/VI/119}, in the hope that BT2
line list will be a vauable tool for astronomers in both spectroscopy
and atmospheric modelling applications.

One of the problems facing the modellers of the atmospheres of cool
stars and brown dwarfs is the disageement between observation and
model. Nevertheless, considerable progress has been made in recent
years in such areas as: convection, molecular abundances, non-LTE
effects and dust (Hauschildt et al. 1999; Tsuji 2002; Alexander et
al. 2003). At the same time, there have been advances in computing the
opacity effects of the various species that are included in the
models.

At the temperature of late M-dwarfs (2,500K), water is by far the most
important contributor to stellar atmospheric opacity, typically
contributing over 60$\%$ of all opacity in the infra-red. It is also
an extremely important contributor to the opacity of L and T brown
dwarfs, although at these lower temperatures (down to 900K), methane
and ammonia play an increased role.

As part of our on-going work in applying BT2 we plan to compare
observed M-dwarf spectra with synthetic spectra produced with the
Phoenix model (Hauschildt, Allard \& Baron 1999) using several water
line lists, including BT2. This is an extension of work already
conducted on modelling oxygen rich cool stars. Jones et al. (2005)
compare the CO 2-0 bands in the 2.297-2.210 $\mu$m region of several
M dwarfs and an L dwarf with opacity calculations using the PS and BT
line lists and conclude that for this particular wavelength range
`While the Partridge-Schwenke line list is a reasonable spectroscopic
match [for the BT line list] at 2,000K, by 4,000K it is missing around
25 $\%$ of the water vapour opacity.'

\section{Acknowledgments}
R.J.Barber has been sponsored in this work by PPARC.\\
Calculations were performed at the UCL Hiperspace computing centre and
we thank its manager, Callum Wright, for his assistance. 
We also thank Svetlana Voronina for performing consistency checks on our energy
level labels.

\label{lastpage}


\begin{thebibliography}{10}
%specifies that there are a double-digit number of entries
\bibitem{AFT03}
Alexander D.R., Ferguson J.W., Tamanai A., Bodnarik J., Allard F.,
Hauschildt P.H., 2003, in Hubeny I., ASP Conf. Series 288.

\bibitem{AHM94}
Allard F., Hauschildt P.H., Miller S., Tennyson J., 1994 ApJ 426 L39.

\bibitem{AHB96}
Allard F., Hauschildt P.H., Baraffe I., Chabrier G., 1996 ApJ 465, L123.

\bibitem{BBA05}
Banerjee D.P.K., Barber R.J. Ashok N.M., Tennyson J., 2005, ApJ 627, L141

\bibitem{BHT02}
Barber R.J., Harris G.J., Tennyson J., 2002, J. Chem. Phys., 117, 11239

\bibitem{BJH_IP}Barber R.J., Jones H.R.A., Hauschildt P., Tennyson J. (in
preparation)

\bibitem{BLA96}Ba\v{c}i\'{c} Z.,Light J.C., 1989,
Annu. Rev. Phys. Chem., 40, 469

\bibitem{Betal96}
Barlow M.J., et al., 1996 A\&A, 315, L241

\bibitem{B96}
Bernath, P.F., 1996, Chem. Soc. Rev., 25, 111

\bibitem{BunkerJensen}
Bunker P.R. \& Jensen P., 2005, 'Fundamentals of Molecular Symmetry', IoP
\bibitem{CFM77}
Camy-Peyret C., Flaud J.-M., Maillard J.-P., Guelachvili G., 1977,
Mol. Phys., 33, 1641

\bibitem{CTN04}
Carr J.S., Tokunaga A.T., Najita J., 2004, ApJ 603, 213

\bibitem{CH87}
Carter S., Handy N.S., 1987 J. Chem. Phys., 87, 4294 

\bibitem{CGA94}
Cernicharo J., Gonz\'{a}lez-Alfonso E.,Alcolea J., Bachiller R., John
D., 1994, ApJ, 432, L59 

\bibitem{CRT69} Cheung A.C., Rank D.M., Townes C.H., Thornton D.D.,
Welch W.J., 1969, Nature, 221, 626

\bibitem{Cetal05}
Coheur P.-F.et al., 2005 J. Chem. Phys., 122, 074307

\bibitem{DMD00}
Dello Russo N., Mumma M.J., DiSanti M.A. Magee-Sauer K., Novak R.,
Rettig T.W., 2000, ICARUS, 143, 324

\bibitem{DSM04}
Dello Russo N., DiSanti M.A., Magee-Sauer K., Gibb E.L., Mumma M.J.,
Barber R.J., Tennyson J., 2004, Icarus, 168, 186

\bibitem{DBD05}
Dello Russo N., Bonev B.P., DiSanti M.A., Mumma M.J., Gibb E.L.,
Magee-Sauer K., Barber R.J., Tennyson J., 2005, ApJ, 621, 537

\bibitem{DGZ05} Dupr\'{e} P., Gherman T., Zobov N.F., Tolchenov R.N.,
Tennyson J., 2005, J. Chem. Phys, 123, 154307

\bibitem{EPAPS}
EPAPS Document No. E-JCPSA6-122-010506
(http://www.aip.org/pubservs/epaps.html).

\bibitem{GMW96}
Gensheimer P.D., Mauersberger R., Wilson T.L., 1996, A\&A, 314, 281

\bibitem{GCB95}
Gonz\'{a}lez-Alfonso E., Cernicharo J., Bachiller R., Fuente A., 1995,
A\&A, 293, L9

\bibitem{HAB99}
Hauschildt P.H., Allard F., Baron E., 1999, ApJ, 512, 377

\bibitem{HB79}
Hinkle K.H., Barnes T.G., 1979, ApJ, 227, 923

\bibitem{JS98}
Jennings D.E., Sada P.V., 1998 Science 279 844

\bibitem{JPV05}
Jones H.R.A., Pavleko Y., Viti S., Barber R.J., Yakovina L.A.,
Pinfield D., Tennyson J., 2005, MNRAS, 358, 105

\bibitem{JJS01}
J{\o}rgensen U.G., Jensen P., S{\o}rensen G.O., Aringer B., 2001,
A\&A, 372, 249 

\bibitem{LMT95}
Lynas-Gray A.E., Miller S.,Tennyson J., 1995 J. Mol. Spec., 169, 458

\bibitem{L71}
Ludvig C.B, 1971, Appl.Opt.,10, 1057
 
\bibitem{MT04}
Miani A., Tennyson J., 2004 J. Chem. Phys., 120, 2732

\bibitem{MTJ94}Miller S. Tennyson J. Jones H.R.A., Longmore A.J.,1994,
in Thejll P., J{\o}rgensen U.G., Proc. IAU Symp. 146,Molecules in the
stellar environment. Springer-Verlag, Berlin

\bibitem{MDD96}
Mumma M.J., DiSanti M.A., Dello Russo N., Fomenkova M., Magee-Sauer
K., Kaminski C.D., Xie D.X., 1996, Science 272 1310

\bibitem{MT98}
Mussa H.M., Tennyson J., 1998, J. Chem. Phys., 109, 10885

\bibitem{PS97}
Partridge H., Schwenke D.W., 1997, J. Chem. Phys., 106, 4618

\bibitem{PZV97b}Polyansky O.L., Zobov N.F., Viti S., Tennyson J., Bernath P.F.,
Wallace L., 1997, J. Mol. Spectrosc., 186, 422

\bibitem{PZV98}
Polyansky O.L., Zobov N.F., Viti S., Tennyson J., 1998, J. Mol. Spec.,
189, 291

\bibitem{RLR02}
Ryde N., Lambert D.L., Richter M.J., Lacy J.H.,2002, ApJ, 580 447 

\bibitem{SP00}
Schwenke D.W., Partridge H., 2000, J. Chem. Phys., 113, 6592

\bibitem{S03}
Schwenke D.W., 2003, J. Chem. Phys., 118, 6898

\bibitem{SSY93}
Shiba H., Sato S., Yamashita T., Kobayashi Y., Takami H., 1993, ApJ
Sup., 89, 299 

\bibitem{SPZ03}
Shirin S.V., Polyansky O.L., Zobov N.F., Barletta P., Tennyson
J., 2003, J. Chem. Phys., 118, 2124

\bibitem{SYG81}
Smith P.L., Yoshino K., Greisinger H.E., Black J.H., 1981, ApJ, 250, 166 

\bibitem{Setal03}
Smith V.V. et al., 2003, ApJ, 599 L107

\bibitem{T88}
Tennyson J., 1992, J. Chem. Soc. Faraday Transactions, 88, 3271

\bibitem{TZW01}
Tennyson J., Zobov. N.F., Williamson R., Polyansky O.L., Bernath P.F.,
2001, J. Phys. Chem. Ref. Data, 30, 735

\bibitem{}
Tennyson J., Barletta P., Kostin M.A., Polyansky O.L. Zobov N.F., 2002
Spectrochimica Acta A, 58, 663

\bibitem{TKB04}
Tennyson J., Kostin M., Barletta P., Harris G.J., Poyansky O.L.,
Ramanlal J., Zobov N.F., 2004, Comp. Phys. Comm., 163, 85

\bibitem{Tetal05}
Tolchenov R.N. et al., 2005, J. Mol. Spec., 233, 68

\bibitem{T01}
Tsuji T., 2001, A\&A, 376, L1

\bibitem{T02}Tsuji T., 2002, ApJ, 575,26L 

\bibitem{VT00}
Vidler M., Tennyson J., 2000, J. Chem Phys. 113, 9766

\bibitem{VTP97}
Viti S., Tennyson J., Polyansky O.L., 1997, MNRAS, 287, 79

\bibitem{V97}
Viti S., 1997, PhD thesis, Univ. London

\bibitem{VBW03}
Volk K., Blum R., Walker G., Puxley P., 2003, IAU Circ. 8188

\bibitem{WBL95}
Wallace L., Bernath P., Livingston W., Hinkle K., Busler J., Guo B.,
Zhang K.-Q., 1995, Science, 268, 1155

\bibitem{WL92}
Wallace L., Livingston W., 1992, An Atlas of a Dark Sunspot Umbral
Spectrum. Nat. Optical Astron. Observatories, Tucson, AZ

\bibitem{ZPT99}
Zobov N.F., Polyansky O.L., Tennyson J., Lotoski J.A., Colarusso P.,
Zhang K-Q., Bernath P.F., 1999, J. Mol. Spectrosc., 193, 188

\bibitem {ZSP06}
Zobov N.F. et al., 2006 J. Mol. Spec. (submitted)

\bibitem{ZPS00} Zobov N.F., Polyansky O.L., Savin V. A., Shirin S.V.,
2000, Atmos. Oceanic Opt.,13 1024

\end{thebibliography}
\end{document}